\documentclass{evn2004}
\usepackage{txfonts}
\usepackage{graphicx}
\begin{document}

\title{Packet Loss in High Data Rate Internet Data Transfer for eVLBI}

\author{R. Spencer, R. Hughes-Jones, A. Mathews and S. O'Toole}

\institute{Dept. of Physics and Astronomy, University of Manchester, Oxford Rd., Manchester M13 9PL, UK}

\authorrunning{Spencer et al.}
\titlerunning{Packet loss in eVLBI}

\abstract{ VLBI is gradually moving to the point where Gbps data rates
are becoming routine. A number of experiments have shown that the
internet can be used at data rates of several hundred Mbps on
production networks. However use of the network is accompanied by
packet loss. The paper discusses the statistics of packet loss as
found by recent tests and investigates the expected effect of packet
loss on correlator performance and signal to noise ratio on eVLBI
observations. The relative merits of UDP versus TCP are also
discussed.  }

\maketitle

\section{Introduction}

Much of the new science with VLBI requires the use of the best
possible sensitivity. EVN is well placed for this since we have some
of the largest radio telescopes in the world. However further
improvements can only come from using wider bandwidths. The current
tape systems are coming to the end of their lives, however there are
two recent advances in technology that are transforming VLBI: the
introduction of the Mk5 disk-based recording system and the use of the
Internet. Experiments in high bandwidth data transfer using the
Internet have taken place over the last 2 years, culminating in the
first ever real-time VLBI image:

\begin{itemize}

\item September 2002: JBO-WSRT fringes were obtained for the iGRID
2002 exhibition in Amsterdam. Westerbork data were pre-recorded on
disk using the PCEVN system and transmitted via SuperJANET4 in
Manchester, over the EC-funded G\'{E}ANT network to Amsterdam and then
via the new SURFnet link to Dwingeloo where it was stored and then
correlated with JBO data recorded on tape. Peak VLBI data transfer
rates of 500 Mb/s were obtained to Amsterdam (Hughes-Jones et al 2003)
\item October 2002 -- July 2003; various small-scale tests were
undertaken and ftp-vlbi implemented at data rates of a few 10's of
Mbps
\item July 2003, WSRT connected at 2.5 Gb/s to Dwingeloo
\item October -- December 2003. Data transmission tests between
Manchester and Dwingeloo achieved more than 900 Mbps -- see this paper
\item November 2003: An international baseline, Onsala (SE) --
Haystack (USA) was used, producing eVLBI fringes only 15 minutes
after observations were made.
\item November 2003: Onsala Space Observatory connected at 1Gb/s.
\item January 2004: First eVLBI image using Jb, On,
Wb. Data recorded on MkV and transmitted over the normal Internet
connection from JBO (using 155 Mbps connection from JBO-Manchester) and over dedicated links from Onsala and WSRT. Data were
then received on MkV systems at JIVE, buffered on disk and played back
into the correlator.
\item April 2004: Real-time fringes On-Wb (no disks) -- with data
streaming directly from the telescopes to the correlator.
\item 28 April 2004: First image from a real-time eVLBI session
involving Jb, On and Wb. No disks were used. The data were
streamed from telescopes directly into the correlator and fringes
obtained immediately. An image of the gravitational lens system
B0218+357 was produced 4 hours after the observations ended.
\end{itemize}

In spite of these successes, it is still unclear how VLBI should make
the best use of the Internet: data rates may be limited by local
conditions in hardware and software, by the local area network, by the
international networks (e.g.  G\'{E}ANT) and not least by the
protocols used. An important parameter is the effect of packet loss on
the data, and this problem is addressed in this paper. Our data rates
are high compared with the average Internet user and so we must be
aware of the possibility of denial of service to others. eVLBI is a
strong driver (along with high energy particle physics, the GRID and
high performance computing) for increases in the available bandwidth
of networks as is evident from the recent favourable discussions with
Research and Education Network providers. It may well be possible to
achieve data rates of several Gbps per telescope in the not too
distant future.

\section{Tests on the Network}

An investigation of the link from the University of Manchester to JIVE in 
Dwingeloo was undertaken as a 4$^{th}$ year MPhys project by a pair of 
undergraduate students (Mathews and O'Toole) in the Autumn term of 2003. The 
link used the SuperJANET4 academic network in the UK to connect to London, 
then G\'{E}ANT to Amsterdam, followed by SURFnet to JIVE in Dwingeloo.

There are two main protocols in common use on the Internet, determined
by software in the sending and receiving machines: Transmission
Control Protocol (TCP/IP) which is used by most ftp systems, and User
Datagram Protocol (UDP) which as the name suggests can be modified to
suit by the user. In both systems data are congregated into packets:
the larger the packets the higher the throughput, limited by the maximum
size of packet that can be accommodated by routers in the link. In our
case this was 1500 bytes, allowing 1472 bytes of user data. Data were
placed on the LAN using 1-Gigabit Ethernet connections. Data rates of
close to 1 Gbps can be achieved provided appropriate network interface
cards and machines are used (Hughes-Jones et al. 2004). In fact the
record at the time of writing (July 2004) stands at 6.6 Gbps on a
Geneva--Los Angeles link using 10-Gigabit cards. TCP/IP produces a
bit-wise correct transfer and tries to be fair for other users.
Packets are checked on arrival and an acknowledgement sent back to the
transmitter. A missing packet is interpreted as congestion and the
transmit rate is halved (Stevens 1993). This can result in a highly
variable transmit rate, but with no missing or corrupted data and
perhaps explains the variable data rates obtained in recent eVLBI
experiments (Parsley priv.  comm.). UDP however will transmit at a
rate determined by the user and the available bandwidth, and has no
acknowledgement. Packets can therefore be lost or out of order with no
effect on the transmit rate, though the receive rate will be less if
packets are lost!

\begin{figure}
  \centering
  \includegraphics[width=6cm,angle=270]{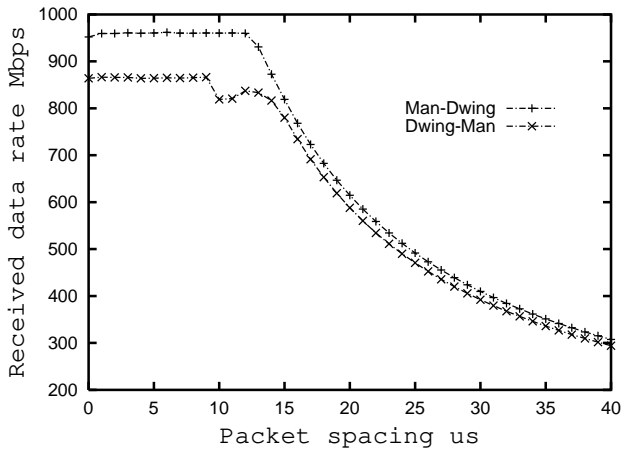} 
 \caption{ Received data rate as a function of packet spacing on the 
Manchester-Dwingeloo link, and in the reverse direction. The two curves show 
that the link is asymmetric, and reflects the different computer power 
available at each end.
         \label{fig:through}
         }
   \end{figure}

\begin{figure}
   \centering
    \includegraphics[width=6cm,angle=270]{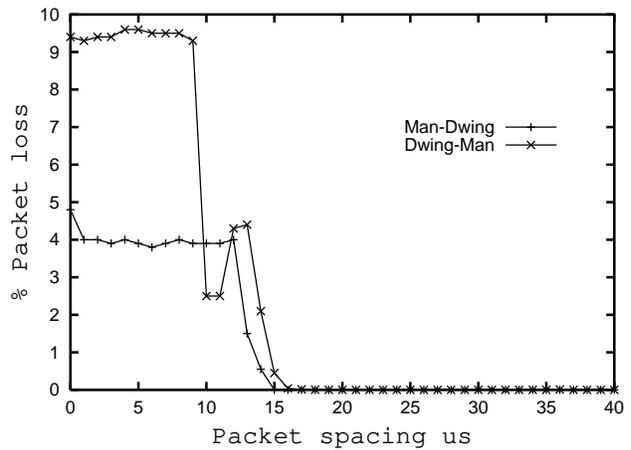} 
   \caption{Packet loss versus packet spacing on the
       Manchester--Dwingeloo link.
   \label{fig:loss}
      }
    \end{figure}

The main aim of the project was to find the distribution of packet
loss in the data transfers and so UDP was chosen. A monitoring program
UDPmon (available from http://www.hep.man.ac.uk/u/rich/) automatically
sends packets and finds data rates and loss as a function of packet
size and inter-packet interval.  Fig.~\ref{fig:through} shows the data
rates achieved using a 2 GHz Xeon machine at Manchester and the 1.2
GHz PIII MkV machine at Dwingeloo for tests made on 11 November
(Man--Dwing) and 13 November (Dwing--Man). Near wire rates were achieved
at maximum, using 10$^{6}$ packets from UDPmon simulating near-continuous
data transfer. The fall off as 1/(packet spacing) occurs because 
the time separating transmission of the packets is dominant over the physical transmission time 
and the
link is waiting for data. The flat part of the curve indicates that
the link (and computers) are limiting the rate. Fig.~\ref{fig:loss}
shows that it is in this flat region that packet loss occurs.

\begin{figure}[htbp]
  \centering
  \includegraphics[width=6cm]{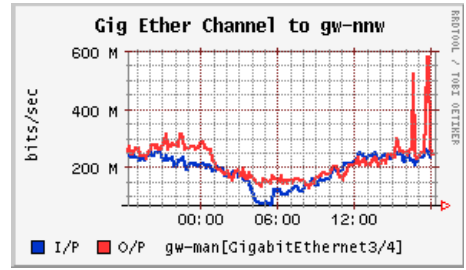}
  \caption{ Traffic to the Net North West router in Manchester showing
  the effect of the tests as the large spikes at around 17:00 hrs.
        \label{fig:traffic}
    }
   \end{figure}

Packet loss can occur due to insufficient processor power or bus
capacity in the end machines, or by congestion in routers on the
link. Fig.~\ref{fig:traffic} shows the traffic on the link to the Net
North West router from Manchester University averaged over 5 minute
intervals.  Our tests, with average data rates of $\sim $400 Mbps,
clearly dominate the traffic, and so congestion occurs when our data
rate reaches 900 Mbps, close to the capacity of one of the two 1
Gigabit Ethernet links used to form the Ether Channel.

\section{Effect of Packet Loss on VLBI data}

Loss of data will cause a decrease in signal to noise (S/N) in VLBI
observations. In normal circumstances S/N will be proportional to
$\sqrt {1-f} $ where $f$ is the fraction of packets lost. However if
the loss of data is sufficient for the correlator to lose
synchronisation, rather more data can be lost. The MkIV Station Unit in the
MkIV correlator checks parity of each 9-bit (8 plus parity) MkIV VLBI
byte. If more than 10{\%} of the bytes per frame are wrong then the
whole frame of 2500 9-bit bytes is rejected.  Luckily the MkIV can
flywheel synchronisation over to the next frame, but obviously if
successive frames are rejected then synchronisation is lost.  This
gives the MkIV systems some resilience to data loss -- a consequence
of having to deal with drop-outs on tape systems.

In an eVLBI system where lost packets can be replaced by random data,
then on average 50{\%} will have the wrong parity. There are
1472x8-bit bytes in a packet and 2500x9 bit bytes in a VLBI data frame
with 32 tracks, giving 2500x9x4/1472=61.14 packets per frame. (The
value quoted in Hughes-Jones et al. (2003) is in error).

On average therefore 0.2x61=12.2 packets need to be lost per frame
before a frame is rejected. Suppose average packet loss per frame is
$a=L/N_{f}$ where $L$ is the number of packets lost per file and
$N_{f}$ is the number of frames in a file. Assuming Poisson statistics,
the probability of $n$ packets being lost in a frame is given by
\[
P_n =\frac{a^ne^{-a}}{n!}
\]
A frame is rejected if more than 12 packets are lost so the number of frames 
$N_{r}$ rejected per file is:
\[
N_r =N_f (1-\sum\limits_{n=0}^{n=12} {P_n )} 
\]

\begin{figure}
 \centering
 \includegraphics[width=9cm]{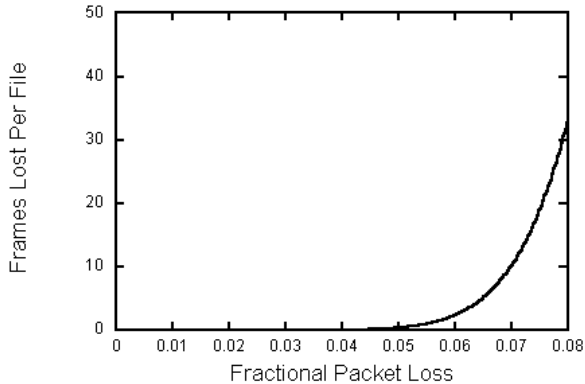} 
 \caption{Number of frames lost per 1.8 Gbyte file versus
  fractional packet loss.
      \label{fig:frames}
      }
   \end{figure}

Fig.~\ref{fig:frames} shows the number of frames lost in a 1.8 Gbyte
file (as used in the iGRID2002 experiment) as a function of the
fractional packet loss $f$ (number of packets lost per file / number
of packets in the file), assuming that missing packets are replaced by
random data. Frame loss is a strong function of packet loss when the
fractional loss exceeds a few \%. More than one frame is rejected in a
file of 1.8 Gbytes (i.e. in 28 seconds of data at 512 Mbps) if there
is more than 5 {\%} packet loss. If missing packets are not replaced
then all missing packets have wrong parity and so one or more frames
are lost if more than 1.5 {\%} packets are lost. Since loss of
synchronisation would mean that the correlator will need some time to
recover, then a figure for packet loss of say less than 2 {\%} is a
useful limiting specification for eVLBI using the Dwingeloo MkV
correlator, since then we would expect either no or rare loss of
synchronisation in either case.

The above calculation is only true if packet loss obeys a Poisson
distributed statistical process, i.e. if successive packet loss is
independent and obeys normal counting statistics. However congestion
is likely to result in correlated bursts in packet loss so long term
correlation is expected. A run on the Manchester-Dwingeloo link during
a time of high packet loss was undertaken to test this idea.

 \begin{figure}
 \centering
 \includegraphics[width=9cm]{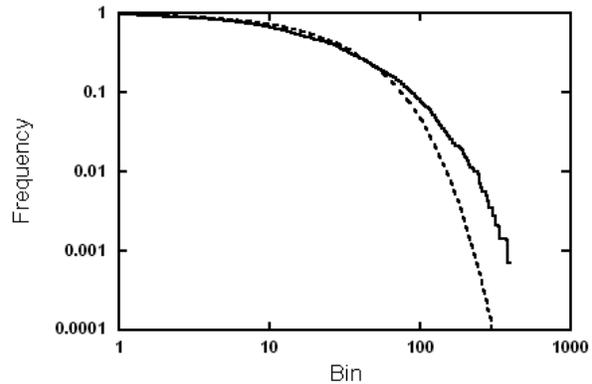} 
 \caption{ Normalised cumulative
   distribution of packet loss (solid curve) and a falling exponential
   fitted to the data (dashed) versus bin number. Each bin is 12 $\mu
   $sec wide.
       \label{fig:cumu}
      }
   \end{figure}

Fig.~\ref{fig:cumu} shows the cumulative distribution of intervals
between lost packets (i.e. $ \int_{t}^{\infty}p(t)\,dt $ where $t$ is
the time interval and $p$ the probability density) found in tests on
4$^{th}$ December 2003 when a total of 1409 packets were lost in a 0.6
sec run.  The distribution of intervals for Poisson process follows a
falling exponential (e.g. Picinbono 1993) and the cumulative
distribution should also be a falling exponential.

 The mean time between lost packets was 424 $\mu $sec, in good
agreement with 394 $\mu $sec found from the fitted exponential.  There
does seem to be an excess of events for bins greater than $\sim$ 100
(1200 $\mu $sec) and a Kolmogorov-Smirnov test shows that this is
significant at the 10 {\%} level. There is therefore some evidence of
long-term effects, however we can conclude that packet loss obeys
Poisson statistics to a reasonable approximation for times up to
around 1 msec.

Long term effects are expected in such data. Analyses of
world-wide-web traffic have shown that the flow is self-similar
(Crovella and Bestavros 1996, Park and Willinger 2000). Traffic occurs
in bursts, and file transmission times have more events with long
transmission times than expected, often obeying a power law
distribution. Packet loss, if related to congestion, is expected to
show similar behaviour, though we have little evidence of a power law
tail in the distribution of packet loss in fig. 4. However these data
were from one run only, more tests covering a wider variety of link
conditions are in progress and might show self-similar effects more clearly.

\section{TCP or UDP?}

The question of which protocol TCP or UDP to use for eVLBI is
important. TCP gives reliable data transfer and is fair to other
users, but could be disadvantageous to eVLBI.  UDP can achieve high
throughput, but could in some circumstances lead to denial of service
for other users, which would not be good politically.

A comparison of throughput can be made by use of models of TCP
behaviour (Padhye et al. 2000). As mentioned above, TCP drops the rate
by a factor of 2 when packet loss occurs. The number of packets in the
window is then incremented by one for each further packet successfully
received. TCP traffic rates therefore follow a saw-tooth pattern. The
average sending rate can only reach 0.75$B$ where $B$ is the available
bandwidth of the link. ${B=WP/R}$ where $W$ is the window size
in packets (i.e. the number of packets in transit), $P$ the packet
size in bytes and $R$ the round trip time (RTT). Allowing for
packet loss and time-outs the model shows that the transmission rate
$T$ is given by
\[
T=\frac{8P}{R\sqrt {\frac{2f}{3}} +6R\sqrt {\frac{3f}{8}} f(1+32f^2)}
\]

    \begin{figure}
 \centering
 \includegraphics[width=9cm]{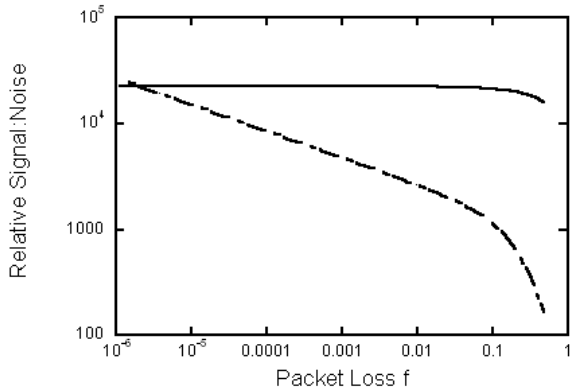}  
  \caption{Comparison of the signal to noise for UDP (upper solid curve) and 
   TCP (lower dashed curve).
       \label{fig:rates}
      }
   \end{figure}

This  formula was  derived  for  the (now  obsolete)  RENO version  of
TCP. An  analysis for more  recent implementations has not yet been made,
but it should still be a good approximation.

The signal to noise ratio (S/N) in one second is proportional to
$\sqrt {B_{W}(1-f)} $ where $B_{W}$ is the bandwidth and $f$ the fraction of
data lost in one second. Fig. ~\ref{fig:rates} shows a comparison of
the S/N for UDP and TCP, when packet loss occurs, assuming that the
rate at which UDP data is sent is 512 Mbps and that the full bandwidth (1
Gbps) is available for TCP. Though TCP will give a higher S/N when
there is no packet loss under these conditions (though under such a
circumstance the UDP rate could also be increased), it can be seen
that the S/N for TCP rapidly deceases when only moderate packet loss
occurs.  UDP is therefore expected to give much better performance for
eVLBI.

An important point missed out so far is the fact that for the standard
TCP stack, the time for TCP to recover its throughput from the loss of
one packet on long distance high bandwidth links can be very large
($=B R^{2}/(2 P)$), i.e. $\sim$ minutes for trans-European links (RTT
$\sim$ 20 ms) and $\sim$ hours for trans-Atlantic links (RTT $\sim$
150 ms). The link may therefore never get into equilibrium as assumed
in the modelling described above, and so TCP data rates may be even
lower. This dramatic variation in TCP throughput will affect the time
that the data from each telescope is presented to the receiving
program. As the path from each telescope will have different packet
loss and RTTs, this implies difficulties in arranging suitable
buffering of the data to maintain presentation of corresponding VLBI
data from the telescopes to the correlator.

\section{Conclusions}

Experiments made on existing networks show that data rates close to
the limiting capacity of 1 Gbit Ethernet are possible on European
networks, at least for short periods, and that rates of 512 Mbps
should not be considered unreasonable. Packet loss should be kept to
less than 2{\%} to avoid loss of synchronisation in the correlator if
UDP is used. Our experiments show that the assumption of Poisson
statistics for the distribution of packet loss is a good
approximation, but further work is needed to clarify expected power
law behaviour at long intervals. UDP gives higher throughput when
packet loss occurs, the resultant decrease in TCP rates has a
devastating effect on signal to noise ratio. Other implementations of
TCP (e.g. High Speed TCP, Fast TCP and TCP Friendly Rate Control) may
offer more optimal solutions, by maintaining high throughput even in
the presence of packet loss. These problems are shortly to be
investigated further by PDRAs working at the University of Manchester.
Useful comments on high data rate transfer can be found on
http://grid.ucl.ac.uk/nfnn.html.

\begin{acknowledgements}

The authors thank Steve Parsley and the staff at JIVE for their valuable 
assistance in running the experiments to Dwingeloo.
\end{acknowledgements}

\noindent
 {\bf References}

\noindent Crovella, M. E., {\&} Bestavros, A., 1996, Proc ACM Sigmetrics, May

\noindent Hughes-Jones, R., Clarke, P., Dalison, S., {\&} Fairey,
G.. 2004, Fut. Gen Comp. Sys., in press

\noindent Hughes-Jones, R., Parsley, S., {\&} Spencer, R., 2003,
Fut. Gen Comp. Sys.  19, 883

\noindent Stevens, W. R., 1993, `TCP/IP Illustrated Vols 1 and 2',
Addison Wesley

\noindent Padhye, J., Firoiu, V., Towsley, D. F., {\&} Kurose, J. F.,
2000 IEEE/ACM Trans Net, 8, 133

\noindent Park, K., {\&} Willinger, W., 2000, Self-Similar Network
Traffic and Performance Evaluation, Wiley

\noindent Picinbono, R., 1993, Random Signals and Systems, Prentice Hall

\end{document}